\documentclass[12pt,preprint]{aastex}
\usepackage{amsmath}

\shortauthors{Zucker et al.}
\shorttitle{New Stellar Structure Near M31}

\begin{document}

\title{
A New Giant Stellar Structure In the Outer Halo of M31}

\author{Daniel B.\ Zucker\altaffilmark{1}, 
Alexei Y.\ Kniazev\altaffilmark{1}, Eric F.\ Bell\altaffilmark{1},   
David Mart\'{\i}nez-Delgado\altaffilmark{1}, Eva K.\ Grebel\altaffilmark{1,2}, Hans-Walter Rix\altaffilmark{1}, 
Constance M.\ Rockosi\altaffilmark{3}, Jon A.\ Holtzman\altaffilmark{4}, Rene A.\ M.\
Walterbos\altaffilmark{4}, \v{Z}eljko Ivezi\'{c}\altaffilmark{5},  J.~Brinkmann\altaffilmark{6}, Howard
Brewington\altaffilmark{6}, Michael Harvanek\altaffilmark{6}, S.\ J.\ Kleinman\altaffilmark{6}, Jurek
Krzesinski\altaffilmark{6}, Don Q.\ Lamb\altaffilmark{7}, Dan Long\altaffilmark{6}, Peter~R.~Newman\altaffilmark{6}, Atsuko
Nitta\altaffilmark{6}, Stephanie A.\ Snedden\altaffilmark{6}}

\altaffiltext{1}{Max-Planck-Institut f\"ur Astronomie,
K\"onigstuhl 17, D-69117 Heidelberg, Germany; \texttt{zucker@mpia.de}}
\altaffiltext{2}{Astronomisches Institut, Universit\"{a}t Basel, Venusstrasse 7,
CH-4102 Binningen, Switzerland}
\altaffiltext{3}{Astronomy Department, University of Washington, Box 351580,
Seattle WA 98195-1580}
\altaffiltext{4}{Department of Astronomy, New Mexico State University, 1320 Frenger Mall,
  Las Cruces NM 88003-8001}
\altaffiltext{5}{Princeton University Observatory
Princeton, NJ 08544}
\altaffiltext{6}{Apache Point Observatory, P.O. Box 59, Sunspot, NM
88349}
\altaffiltext{7}{Astronomy and Astrophysics Department, University of
  Chicago, 5640 S. Ellis Ave, Chicago, IL 60637}

\begin{abstract}

The Sloan Digital Sky Survey has revealed an overdensity of luminous
red giant stars $\sim
3\degr$ (40 projected kpc) to the northeast of M31, which we have called Andromeda NE. The
line-of-sight distance to Andromeda NE is within $\sim 50$\ kpc of M31; Andromeda
NE is not a physically unrelated projection. Andromeda NE has a $g$-band
absolute magnitude of $\sim -11.6$ and central surface brightness of $\sim
29$ mag\,arcsec$^{-2}$, making it nearly two orders of magnitude more
diffuse than any known Local Group dwarf galaxy at that luminosity. Based
on its distance and morphology, Andromeda NE is likely undergoing tidal
disruption. Andromeda NE's red giant branch color is unlike that of M31's
present-day outer disk or the stellar stream reported by Ibata et al.
(2001), arguing against a direct link between Andromeda NE and these
structures. However, Andromeda NE has a red giant branch color similar
to that of the G1
clump; it is possible that these structures are both material torn off of
M31's disk in the distant past, or that these are both part of one ancient
stellar stream.
\end{abstract}


\section{Introduction}

The study of dwarf galaxies offers special insight into the physics
of galaxy formation and evolution.
There are many fewer dwarf galaxies
than predicted by a direct scaling of the dark matter halo
mass function \citep[e.g.,][]{whit91}.  A number of theoretical
explanations for this have been explored: feedback from supernovae
leading to disruption of dwarf galaxies \citep[e.g.,][]{deke86}
or strong reduction of their luminous/dark matter ratio 
\citep[e.g.,][]{haya03}, reduced substructure in warm dark matter haloes
\citep[e.g.,][]{moor00}, enhanced satellite disruption and reionization
\citep[e.g.,][]{bull01}, or suppression of dwarf galaxy
formation through photoionization \citep[e.g.,][]{some02}.  In this context, 
the dwarf galaxy content of the Local Group plays a special role.
The luminosity function can be probed down to very faint limits, 
where the discrepancies with the models are largest \citep{moor99}.  
Furthermore, the detailed star formation histories
of the dwarfs can be estimated through analysis of the
colors and magnitudes of their resolved stellar populations.
Coupled with the exquisitely deep surface brightness limits
achievable through analysis of resolved stellar populations, 
one can explore the physical processes which shape dwarf galaxy 
evolution directly, such as ram pressure stripping 
\citep[e.g.,][]{greb03} or tidal 
disruption by the giant galaxies in the Local Group \citep[e.g.,][]{ibat94}. 

Nevertheless, constructing a 
complete census of faint stellar structures (dwarf galaxies, 
disrupted dwarfs, stellar streams) has proved challenging, owing to the twin 
constraints of imaging depth and large area coverage 
\citep[see, e.g.,][]{arma99,ferg02}.
In this letter, we report the discovery of a faint, extended
stellar feature in the outer halo of M31, 
using a 
special scan from the Sloan Digital Sky Survey 
which reaches depths sufficient 
to detect stars at the tip of the red
giant branch (TRGB).
We use these data to estimate the feature's physical properties and to try to determine 
its true nature. We assume
a distance to M31 of 760\,kpc \citep{vand99}.

\section{Observations and Data Analysis}\label{obsdatatxt} 

The Sloan Digital Sky Survey (SDSS) \citep{york00} is an
imaging and spectroscopic survey that will eventually cover about $1/4$
of the sky.
Drift-scan imaging from the SDSS 2.5m telescope 
in the five SDSS bandpasses ($u,g,r,i,z$)
\citep{fuku96,gunn98,hogg01} is processed through data reduction pipelines
to measure photometric and astrometric properties
\citep{lupt02, stou02, smit02, pier03} and
to identify targets for spectroscopy.
M31 was observed on 5 October 2002, during time when 
primary survey areas were inaccessible. 
The data were obtained in two overlapping strips oriented along the 
major axis of M31, covering an area of $\sim 18\degr \times 2\fdg5$.
Conditions were photometric, but the seeing varied between
$1 \arcsec - 2 \arcsec$ FWHM in $r$-band. 
The raw data were processed and catalogs constructed using the same 
pipeline as Data Release 1 (DR1) \citep{abaz03}.  
We selected only objects within the coordinate 
range assigned to each strip, thus eliminating duplicate object 
detections. In the
following, all references
to dereddening and conversion from SDSS magnitudes to
$V,I$ magnitudes (for literature comparison) utilize \citet{schl98} and
\citet{smit02}, respectively. 

\begin{figure*}[th]
\epsscale{1.0}
\plotone{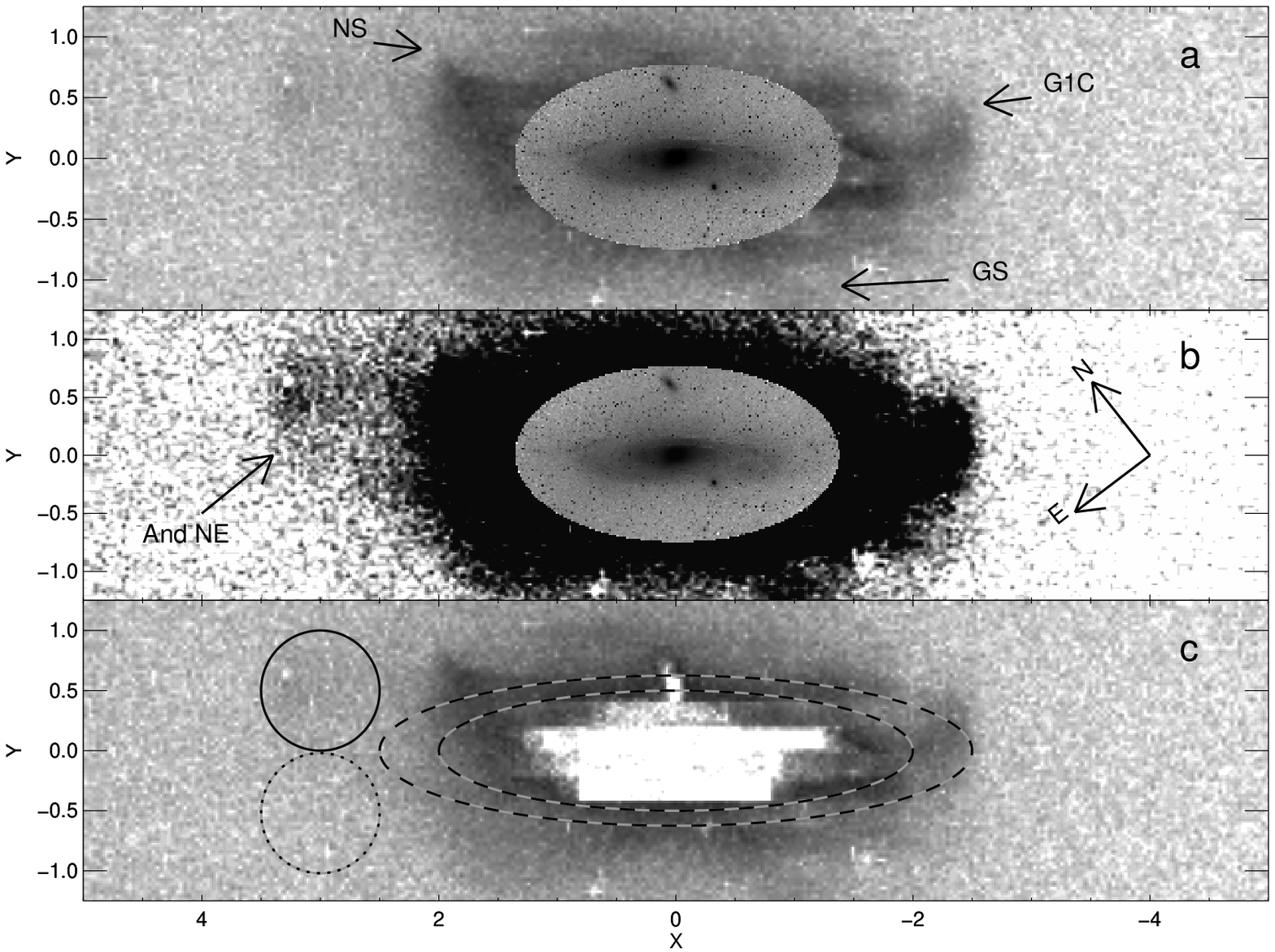}
\caption{\label{stardistfig} M31 Stellar Substructure in the SDSS
Data: a) The spatial distribution of all stars with $21~< ~i~\leq~ 22$, with an inset optical image of M31 to scale. 
The arrows indicate major halo substructures shown in \citet{ferg02}: the northern spur (NS), the giant stream (GS) and
the G1 clump (G1C). 
b) Same as a), but with the grayscale adjusted to emphasize faint
details. The arrow shows the location of Andromeda NE. The orientation
of the scan is as indicated. c) Same as a), without the inset image,
indicating the region used for the M31 sample (between the ellipses)
and that used for And NE (the solid circle); the dotted circle shows the control field used in our analysis. The data have been binned $2\arcmin \times 2\arcmin$. X and Y are in arcdegrees from the center of M31 ($00^{\rm h}42^{\rm m}44\fs3 +41\degr16\arcmin08\farcs5$ J2000), along the major and minor axes, respectively. The inset image of M31 is from Bill Schoening, Vanessa Harvey/REU program/NOAO/AURA/NSF.
}
\end{figure*}

The spatial density of SDSS-detected stars with $21 <  i \leq 22$ is shown 
in Fig.~\ref{stardistfig}. The central regions of M31, its disk, and 
its bright satellites M32 and NGC205 are too crowded for the 
standard SDSS photometric pipeline. The slight horizontal ``striping'' is the result of 
seeing variations; under better seeing conditions, 
more stars were detected. This effect is most pronounced at faint magnitudes.
Complex stellar substructure is evident in the outer disk and halo of
M31 (Fig. \ref{stardistfig}a). Essentially all of the morphological
features visible in Figures 2 and 3 of \citet{ferg02} can also be seen
in the SDSS data, although the giant stellar stream  to the southeast \citep{ibat01,ferg02,mcco03} is truncated by the limited width ($2\fdg 5$) of the SDSS scan.

Adjusting the grayscale in Fig. \ref{stardistfig}a to enhance faint details, we obtain Fig. \ref{stardistfig}b, in which an overdensity around $(X,Y) \sim (0.5,3.1)$, or $00^{\rm h}52\fm0 +44\degr06\arcmin$ (J2000), becomes apparent. The seeing, rather fortuitously, was almost constant in the vicinity of this overdensity, minimizing the striping in this area.
The stellar overdensity 
has its highest concentration at a angular separation of $\sim 3\fdg3$
from the center of M31, and although it is adjacent to the area mapped
by \citet{ferg02}, it is not covered by the data shown in that work.\footnote[1]{Subsequent to initial
  submission of this work, \citet{lewi04} presented a map which
  clearly shows this feature, although the authors did not comment on
  its significance.}

\begin{figure}[th]
\epsscale{0.70}
\plotone{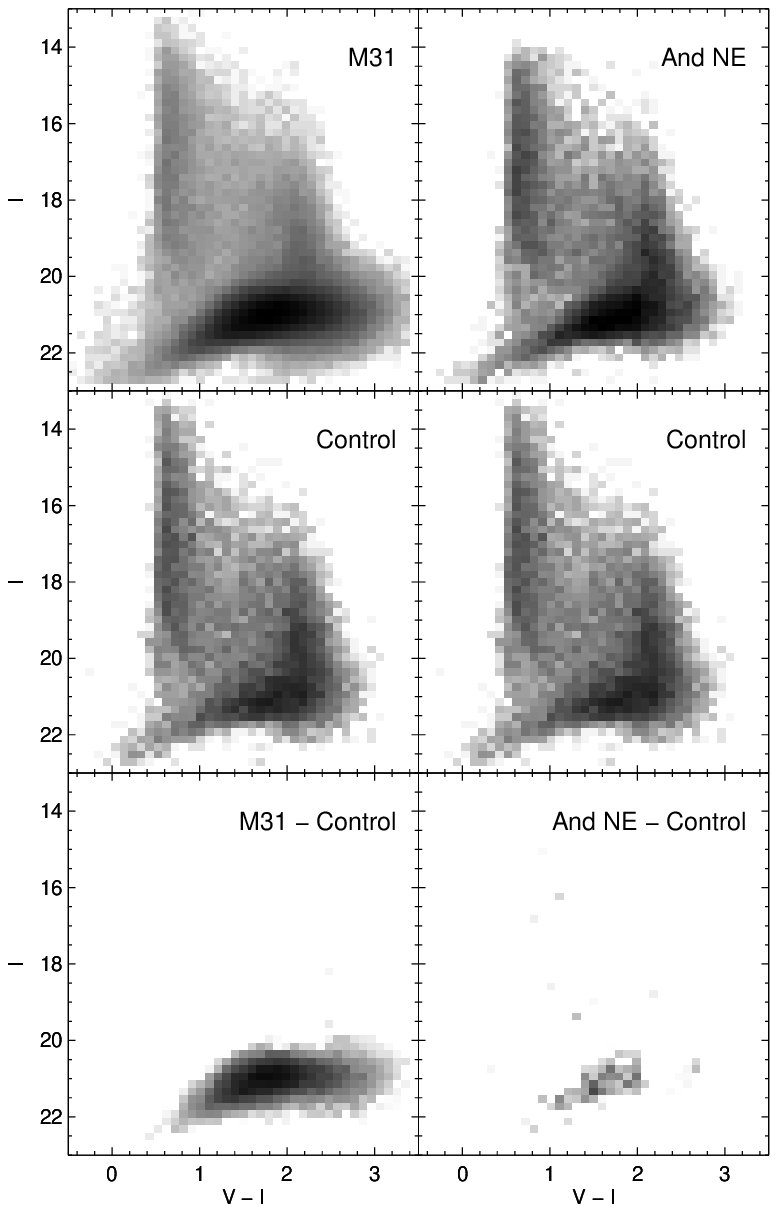}
\caption{\label{cmdfig} Hess diagrams  of
M31 and And NE: {\it Top left:} Hess diagram of
all stars within an inner halo ellipse of M31. 
{\it Middle left:} Hess diagram of all stars within the 
 control field (see Fig. \ref{stardistfig}c).
{\it Bottom left:} The difference of the M31 Hess diagram and the scaled control field Hess diagram, divided by the square root of the sum of the Hess diagrams.
{\it Top right:} Hess diagram of all stars within a $0\fdg5$ radius circle centered on And NE. 
{\it Middle right:} Hess diagram of all stars within the 
 control field.
{\it Bottom right:} The difference of the And NE
Hess diagram and the scaled control field Hess diagram, divided by the square root of
the sum of the Hess diagrams. 
The data were dereddened and transformed to $V,I$, binned by $0.2$ mag
in $I$ and $0.1$ mag in $(V - I)$, and are shown with logarithmic
grayscales.}
\end{figure}

In order to place constraints on the nature and distance of this new           
stellar overdensity, we dereddened the data, transformed $g,r,i$
photometry to $V,I$ magnitudes, and constructed color-magnitude diagrams (CMDs) of a
region in the outskirts of M31, of the new feature, and of a nearby control
field (Fig. \ref{stardistfig}c). The left side of Fig. \ref{cmdfig}
shows the Hess diagrams of the M31 periphery (between the two ellipses
in Fig. \ref{stardistfig}), the control field, and the
foreground-corrected M31 Hess diagram. The right side of
Fig. \ref{cmdfig} shows the same for the new, diffuse stellar feature.
The new stellar feature has a Hess diagram which is qualitatively similar to
that of M31, placing it roughly at the distance of M31, and firmly
establishing its extragalactic nature.

\begin{figure}[th]
\epsscale{1.0}
\plotone{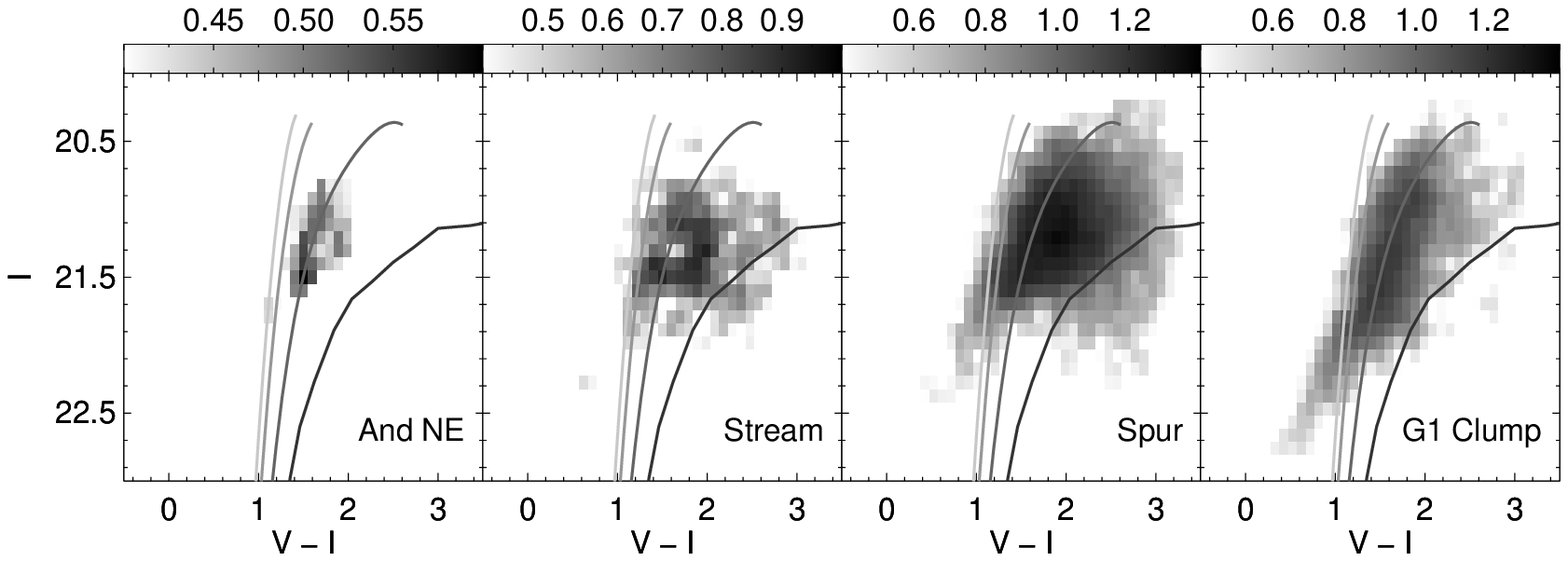}
\caption{\label{rgbcolorfig} RGB Colors of Stellar Structures from
  SDSS Data: {\it Left}: Hess diagram of And NE, minus an
  appropriately scaled control field Hess diagram, divided by the
  square root of the sum of the two Hess diagrams (to estimate S/N). {\it Left middle}:
  Same as ({\it left}), but for a field in the Giant Stream
  \citep{ibat01}. {\it Right middle}: Same as ({\it left}), but for a
  field in the Northern Spur. {\it Right}: Same as ({\it left}), but for a
  field in the G1 Clump.
The data were dereddened and transformed to $V,I$ magnitudes, binned
by $0.1$ mag in $I$ and $V - I$, and smoothed with a Gaussian filter. The
grayscale bar at the top of each panel indicates the logarithmic range in
estimated S/N shown for that Hess diagram.
Fiducial sequences are overplotted for Galactic globular clusters with metallicities of (left to right) [Fe/H] = $-2.2$ (M15),
$-1.6$ (M2), $-0.7$ (47~Tuc), and $-0.3$ (NGC~6553) \citep{daco90,
  saga99}, shifted to the adopted M31 distance modulus of 24.4;  for
NGC~6553, $(m - M)_0 = 13.7$ and $E(V - I) = 0.95$ were assumed \citep{saga99}.}
\end{figure}

\section{A New Stellar Structure: Andromeda NE} 

The stellar overdensity revealed in Fig. \ref{stardistfig}b, which
we will call Andromeda NE (And NE)
 in what follows,
covers more than half a square degree.
To minimize foreground star contamination in such an extended, faint stellar
feature, we selected stars in those Hess diagram bins from the bottom
left panel of Fig. \ref{cmdfig} with S/N $\geq 3.0$ (i.e., those likely to be at the distance of M31)
for further analysis \citep[cf.][]{oden01}.
The SDSS data only include stars near the TRGB (2.6 arcmin$^{-2}$ at
the densest part of the feature), and so to estimate And NE's
peak surface brightness and total luminosity we carefully accounted
for this effect. Following the method of \citet{knia04}, we
measured the total luminosity of all selected stars in And NE in circular
apertures, minus an average stellar background level determined from
nearby external fields. Analyzing SDSS data for the Pegasus dwarf irregular
(DDO216), a galaxy at a distance comparable to that of M31, we
fitted surface brightness profiles to both stars and to the
total integrated light in its outer regions, determining offsets of
$\sim 2.8, 2.5, {\rm and~} 2.0$ mag in $g,r, {\rm and~} i$ between the
`resolved-source' and total surface brightnesses \citep[an $i$-band correction of 1.8 mag was derived from 
the $i$-band stellar luminosity function of the 
Draco dwarf spheroidal given in][confirming that these estimates are
reasonable for a range of likely stellar populations]{oden01}.
We then applied these corrections to the And NE measurements,
obtaining central surface brightnesses and total magnitudes of 
$29\pm0.4,28.3\pm0.3, {\rm and~} 27.2\pm0.3$ mag arcsec$^{-2}$, and
$12\fm8, 11\fm4, {\rm and~} 11\fm0$ in $g$, $r$, and $i$-band, respectively.
If we assume that And NE is at approximately the same distance
as M31 (justified in the next paragraph), and correct for a median foreground
extinction of $A_{g} \sim 0.32~ (A_{V} \sim 0.28)$, it has $M_{g} \sim -11.9~ (M_{V}
\sim -12.7)$;
Local Group dwarfs of comparable luminosity have $V$-band
surface brightnesses nearly two orders of magnitude higher than
And NE \citep[see, e.g.,][]{greb03}.  
It is clear that And NE, if it is a galaxy, 
is a highly unusual one.

We estimated the distance to And NE from the
position of the TRGB \citep[cf.][]{lee93} in And NE.
We generated cumulative dereddened $I$-band luminosity functions for
stars within the two circular regions shown in Fig. \ref{stardistfig}c,
and subtracted the control field luminosity function from that of And
NE. The field-star-subtracted luminosity function rises noticeably at
$I \sim 20.4$; for RGB metallicities in the range $-2.2 \leq {\rm
  [Fe/H]} \leq -0.7$, $M_{I,TRGB} \sim -4.0\pm0.1$ \citep{daco90,sala98}, yielding a
distance modulus of $(m - M)_0 \sim 24.4\pm0.15\, (760^{+55}_{-50}\, {\rm
  kpc})$ for And NE. Thus we conclude that And NE is at
approximately the same distance as M31. However, the precision of this
result is limited by the small number of stars at the TRGB in And NE
and the possibility that And NE is extended along the
line of sight; significant improvement will likely only be possible using, 
e.g., more numerous red clump or horizontal branch stars.

What is And NE? To give more insight into its nature, we compared the Hess
diagrams for the red giant branch (RGB) in And NE and in three other
prominent stellar features: the giant stream, the northern spur,
and the G1 clump. For each field, we dereddened the SDSS data,
transformed $g,r,i$ stellar magnitudes to $V$ and $I$, subtracted an
appropriately scaled control field Hess diagram, and finally divided by the square root
of the sum of the target and control field Hess diagrams to estimate S/N
in each bin. Fig. \ref{rgbcolorfig} shows the results, with overplotted RGB
fiducials for Galactic globular clusters spanning a wide range of
metallicities. And NE has a ``blue" RGB \citep[following the nomenclature of][]{ferg02}, unlike the ``red" RGBs shown by the
northern spur and the giant stream, from
which we infer that And NE has a lower mean metallicity. For this reason,
we believe it unlikely that And NE is an extension of either the nearby
northern spur or the giant stream.

Interestingly, And NE and the G1 clump have very similar RGB morphologies,
despite the large differences in their stellar densities and projected
positions (70 kpc on opposite sides of M31's disk). Moreover, And
NE appears to have a faint extension towards the disk of M31.
Taken together, these data suggest two possibilities for And NE. The first
is that the G1 clump and And NE are both material torn from M31's
disk by an ancient interaction; this possibility is supported by the
disk-like kinematics of RGB stars in the G1 clump \citep{reit04}, but is
argued against by the strong difference in RGB morphology between 
M31 disk stars and And NE (M31's disk
appears more metal-enriched than And NE or the G1 clump; see
Figs. \ref{cmdfig} and \ref{rgbcolorfig}). Secondly, it is possible that And NE and the G1 clump
are both projections along the orbit of an ancient, diffuse stellar
stream roughly in the plane of M31's disk. The morphology of both
features and the lack of similarity between disk and And NE stellar
populations support this possibility. The disk-like kinematics
of stars in the G1 clump would appear to argue against this scenario,
but such a coincidence is not impossible; the Monoceros stream around the disk of the Milky
Way \citep{newb02,yann03} exhibits prograde, disk-like kinematics
(e.g., Penarrubia et al., in preparation). 

Having discussed possible stellar stream and stellar debris scenarios
for And NE, we turn to the issue of whether And NE is 
a bound, self-gravitating entity (i.e., a new satellite).
We can calculate what mass-to-light (M/L) ratio 
would be required for it to be gravitationally bound 
by using an approximation for tidal disruption for a cluster 
of particles in a circular orbit: $3 M_{\rm M31}/D_{\rm M31}^3 >
M_{\rm And\,NE}/R_{\rm And\,NE}^3$, where $M_{\rm M31}$ and 
$M_{\rm And\,NE}$ are the total masses of M31 and And NE, 
$D_{\rm M31}$ is the distance of And NE from M31, and 
$R_{\rm And\,NE}$ is the radius of And NE.  The masses are 
unknown; however, we can calculate the total M/L of And NE 
that would be required to bind And NE, in units of M31's 
M/L at radii $\ga 40$ kpc.  
The integrated magnitude of And NE is $V_{\rm And\,NE} \sim 12.0$, and using $V_{\rm M31} = 3.44$ from 
the RC3 catalog, and assuming that M31 and And NE have the same
reddening and lie
at the same distance (therefore $R_{\rm And\,NE} = 6.6$\,kpc, 
$D_{\rm M31} = 43$\,kpc), $(M/L)_{\rm And\,NE} \sim 30 (M/L)_{\rm M31}$.
This is high enough to argue that, if And NE is at the same distance as M31, it is likely undergoing tidal disruption.
Yet, And NE may be rather more distant from M31: if instead
we assume that And NE is $\pm 0.15$\,mag more or less distant
(the error in distance determination), one obtains  
$(M/L)_{\rm And\,NE} \sim 6 (M/L)_{\rm M31}$. While this latter value
still suggests that And NE is being tidally disrupted, accurate
velocity and distance data will be required to firmly resolve the question of
 whether And NE is bound or unbound.

\section{Conclusions}

SDSS has revealed an overdensity of red giant stars 
$\sim 3\degr$ to the NE of the Andromeda galaxy; 
Andromeda NE is within $0\fm15$ in distance modulus ($\sim 50$ kpc) of
M31 and is thus not an unrelated foreground or background object.
Its extrapolated total $g$-band magnitude is 
$\sim 12.8$\, ($M_g = -11.6$ or $5\times 10^6 L_{\odot}$
at 760\,kpc), and its radius, although not well-defined, is $\sim
30\arcmin$ (6 kpc). And NE is very diffuse and is probably
undergoing tidal disruption, based on its distance from M31 and its
morphology. And NE possesses a ``blue" RGB, unlike M31's
northern spur, the giant stream, or the outer parts of the
present-day M31 disk; yet And NE's stellar population is similar to the G1
clump, suggesting perhaps a link with this structure. It is possible that
it is part of an ancient diffuse stellar stream, or perhaps material
torn from M31's disk in the distant past. Spectroscopic and
deep, high spatial-resolution data will be required to further
elucidate the nature of this faint, enigmatic stellar feature.

\acknowledgements
DBZ and EFB respectively acknowledge support from an NSF International
Postdoctoral Fellowship and from the European
Community's Human Potential Program contract
HPRN-CT-2002-00316, SISCO.

Funding for the creation and distribution of the SDSS Archive has been
provided by the Alfred P. Sloan Foundation, the Participating
Institutions, the National Aeronautics and Space Administration, the
National Science Foundation, the U.S. Department of Energy, the
Japanese Monbukagakusho, and the Max Planck Society.  The SDSS Web
site is (\texttt{http://www.sdss.org/}).

The SDSS is managed by the Astrophysical Research Consortium (ARC) for the
Participating Institutions. The Participating Institutions are The University
of Chicago, Fermilab, the Institute for Advanced Study, the Japan Participation
Group, The Johns Hopkins University, Los Alamos National Laboratory, the
Max-Planck-Institute for Astronomy (MPIA), the Max-Planck-Institute for
Astrophysics (MPA), New Mexico State University, University of Pittsburgh,
Princeton University, the United States Naval Observatory, and
the University of Washington.

\end{document}